\documentclass[aps,prl,reprint,superscriptaddress,amsmath,amssymb,letterpaper]{revtex4-1}
\usepackage{mathtools,ulem}
\usepackage{graphicx,subfigure,float,dcolumn}
\usepackage{bm,color,xcolor,calc}
\usepackage{tikz,tikz-3dplot,pgfplots,pgfplotstable}
\usetikzlibrary{arrows,arrows.meta,positioning,shapes,calc,3d,fadings,matrix,decorations.pathreplacing}
\usepackage[linktoc=all]{hyperref}
\usepackage{url,titlesec,titletoc}
\usepackage{pagecolor}
\hypersetup{
	colorlinks,
	citecolor=blue,
	filecolor=black,
	linkcolor=purple,
	urlcolor=cyan
}
\pgfplotsset{compat=1.15}
\newcommand{\vecb}[1]{\boldsymbol{#1}}

\newcommand{\cycsum}[0]{\sum\limits_{\mathrm{cyc}}}
\begin{document}
	\title{Monopole topological resonators}
	
	\author{Hengbin Cheng}
	\affiliation{Institute of Physics, Chinese Academy of Sciences/Beijing National Laboratory for Condensed Matter Physics, Beijing 100190, China}
	\affiliation{School of Physical Sciences, University of Chinese Academy of Sciences, Beijing 100049,China}
	
	\author{Jingyu Yang}
	\affiliation{Institute of Physics, Chinese Academy of Sciences/Beijing National Laboratory for Condensed Matter Physics, Beijing 100190, China}
	\affiliation{School of Physical Sciences, University of Chinese Academy of Sciences, Beijing 100049,China}
	
	\author{Zhong Wang}
	\affiliation{Institute for Advanced Study, Tsinghua University, Beijing 100084, China}
	
	\author{Ling Lu}
	\email{linglu@iphy.ac.cn}
	\affiliation{Institute of Physics, Chinese Academy of Sciences/Beijing National Laboratory for Condensed Matter Physics, Beijing 100190, China}
	\affiliation{Songshan Lake Materials Laboratory, Dongguan, Guangdong 523808, China.}
	
	\begin{abstract}
		Among the many far-reaching consequences of the potential existence of a magnetic monopole, it induces topological zero modes in the Dirac equation, which were derived by Jackiw and Rebbi 46 years ago and have been elusive ever since.
		Here, we show that the monopole and multi-monopole solutions can be constructed in the band theory by coupling the three-dimensional Dirac points in hedgehog spatial configurations through Dirac-mass engineering.
		We then experimentally demonstrate such a monopole bound state in a structurally-modulated acoustic crystal as a cavity device.
		These monopole resonators not only support an arbitrary number of degenerate mid-gap modes, but also offer the optimal single-mode behavior possible --- whose modal spacing is inversely proportional to the cubic root of the modal volume.
		Our work completes the kink-vortex-monopole trilogy of zero modes and provides the largest free spectral range for sizable resonators.
	\end{abstract}
	
	\maketitle
	Topological physics can be largely understood through the celebrated Dirac equations~\cite{Shen-2017-book,Jackiw-2012-fractional}, where the gapped massive solutions correspond to the topological insulators in the bulk and the mid-gap zero-energy solutions correspond to the gapless states at the interfaces.
	The simplest zero mode is the one-dimensional~(1D) kink solution obtained by Jackiw and Rebbi~\cite{Jackiw-Rebbi-1976}, in which the Dirac mass forms a domain wall in real space --- the topological defect in 1D. The kink solution has materialized in polyacetylene whose lattice theory is the famed Su-Schrieffer–Heeger model~\cite{SSH-1979-solitons}.
	The 2D topological defect, of Dirac masses, is a vortex whose zero-mode solution was solved by Jackiw and Rossi~\cite{Jackiw-Rossi-1981}. 
	These vortex zero modes are proposed both in graphene with ${\rm Kekul\acute{e}}$ distortions~\cite{Hou-2007-electron} and in the vortex cores between the superconductor and topological insulator~\cite{Fu-2008-superconducting}. The graphene model has recently led to experimental realizations of the vortex mid-gap modes in a number of phononic~\cite{	Gao-2019-majorana,Chen-2019-mechanical,Ma-2021-nanomechanical} and photonic~\cite{Menssen-2020-photonic,Noh-2020-Braiding,Gao-2020-Dirac} lattices.
	In 3D, however, the dimension of our real world in which the Dirac equation was originally written, no progress has been made towards the physical discovery of  the monopole zero mode since its analytical solution was first known in 1976~\cite{Jackiw-Rebbi-1976}.
	In this work, we theoretically propose and experimentally demonstrate this monopole mode in a Dirac crystal using hedgehog modulations as a 3D topological resonator.
	
	The monopole resonator exhibits the best single-mode behaviour ever, i.e., the monopole mode has the largest free-spectral-range~(FSR) among all known resonators in the large cavity limit.
	As discussed in Ref.~\cite{Gao-2020-Dirac}, the mid-gap cavities, of low dimensions, have been established as the ideal robust designs for single-mode selection in semiconductor lasers.
	The modal spectrum of mainstream diode laser products --- powering the Internet communications and cell-phone unlocking --- is identical to that of the 1D kink zero modes, whose FSR is inversely proportional to the mode volume.
	Very recently, the 2D vortex zero mode inspired the invention of the high-brightness topological-cavity surface-emitting laser~\cite{Yang-2022-topological}, whose FSR is inversely proportional to the square root of the mode volume.
	Now, our 3D monopole resonator, whose FSR is inversely proportional to the cubic root of its mode volume, represents the optimal design for the large-volume single-mode operation.
	
	\begin{figure*}[t]
		\includegraphics[scale=1.0]{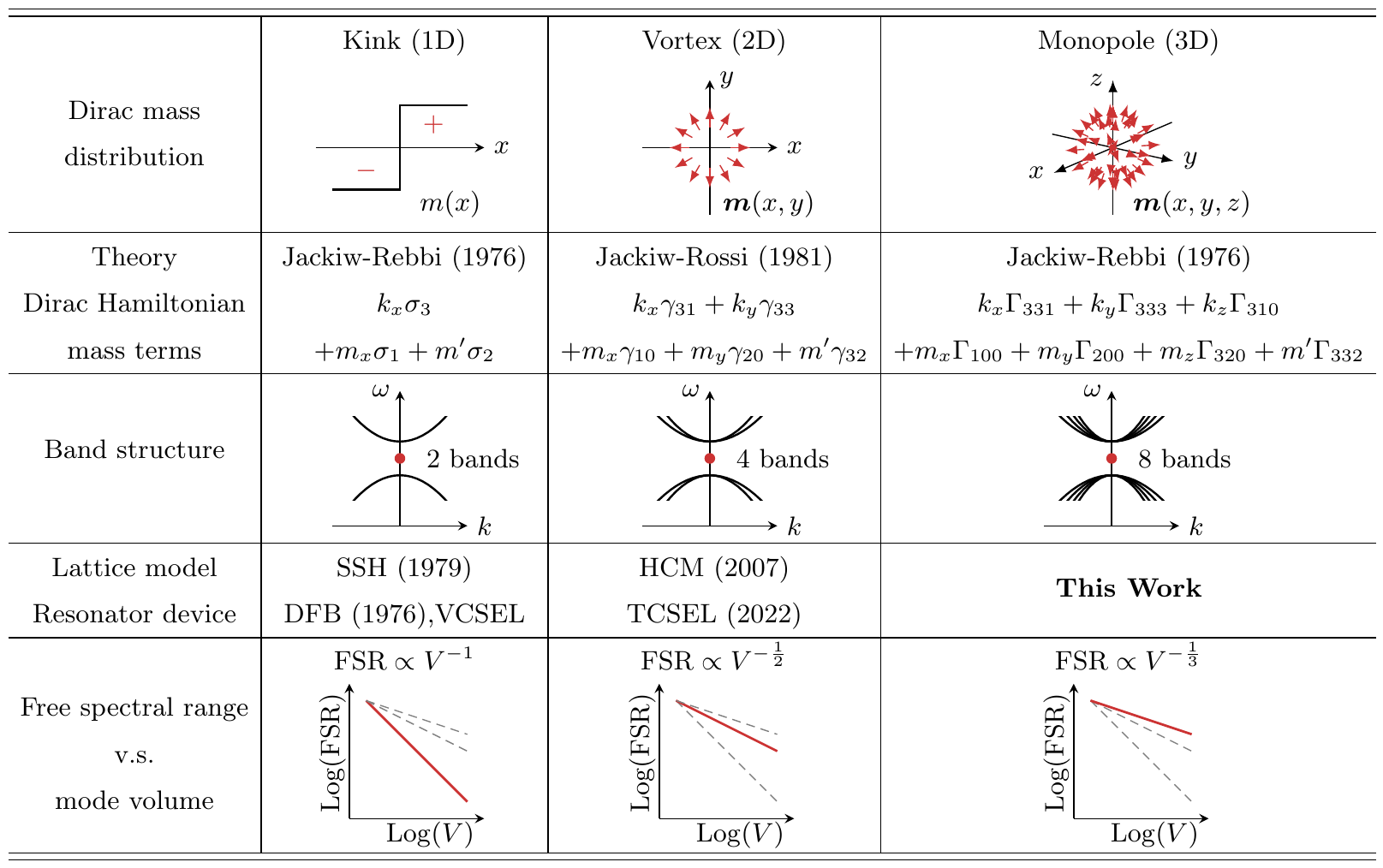} 
		\caption{
			Comparison of topological point defects of the Dirac masses in each dimension.
			$\sigma_i$ are the Pauli matrices, $\gamma_{ij}=\sigma_i\otimes\sigma_j$ are the gamma matrices, and $\Gamma_{ijk}=\sigma_i\otimes\sigma_j\otimes\sigma_k$, where $\otimes$ is the Kronecker product.
			SSH: Su-Schrieffer–Heeger~\cite{SSH-1979-solitons},
			HCM: Hou-Chamon-Mudry~\cite{Hou-2007-electron},
			DFB: distributed feedback~\cite{Haus1976Antisymmetric},
			VCSEL: vertical-cavity surface-emitting laser~\cite{padullaparthi2021vcsel},
			TCSEL: topological-cavity surface-emitting laser~\cite{Yang-2022-topological}.}
		\label{fig:symmary}
	\end{figure*}
	\section*{Jackiw-Rebbi monopole mode}
	After the proposals of the Dirac monopole in Maxwell's equations~\cite{Dirac-1931-quantised} and the 't Hooft-Polyakov monopole in Yang-Mills equations~\cite{Hooft-1974-magnetic,Polyakov-1974-particle}, Jackiw and Rebbi found a zero-mode solution bound to the magnetic monopole in the Dirac equation~\cite{Jackiw-2012-fractional}.
	The monopole zero mode arises at a 3D topological point defect of three independant mass terms in an 8-by-8 extended Dirac equation shown in Fig.~\ref{fig:symmary}. 
	There has to be a minimal of eight bands to allow three mass terms, for the hedgehog construction without breaking the time-reversal symmetry.
	
	We first generalize the Jackiw-Rebbi single-monopole mass profile into the multi-monopole case of arbitrary charges.
	Although the mass distribution can be chosen to be spherically symmetric for a single monopole, it cannot for monopoles of charge greater than one~\cite{Weinberg-1976-nonexistence}.
	Our choice of the spatial mapping of the Dirac-mass vectors~$\vecb{\hat{m}}(\vecb{r})=[\hat{m_x}(\vecb{r}),\hat{m_y}(\vecb{r}),\hat{m_z}(\vecb{r})]$ is expressed as
	\begin{equation}
		\label{eq:monopole}\begin{aligned}
			\hat{m_x}(\vecb{r})&=\sin(W_\theta \theta)\cos(W_\phi \phi)\\
			\hat{m_y}(\vecb{r})&=|\sin(W_\theta \theta)| \sin(W_\phi \phi)\\
			\hat{m_z}(\vecb{r})&=\cos(W_\theta \theta)\\
			W =\frac{1}{8\pi}&\int_A dA \epsilon^{\theta\phi} \epsilon^{xyz} \hat{m_x} \partial_\theta \hat{m_y} \partial_\phi \hat{m_z}=W_\theta W_\phi
	\end{aligned} \end{equation}
	where $\theta$ and $\phi$ are the polar and azimuthal angles in the spherical coordinate; $W_\theta$ and $W_\phi$ are the corresponding winding numbers.
	$W$ is the monopole charge, measuring how many times the 3-component Dirac-mass vectors wraps around a closed surface~($A$) enclosing the monopole~\cite{Fukui-2010-topological}.
	The absolute value in $\hat{m_y}$, or in $\hat{m_x}$, ensured that the 3D wrapping number~($W$) is expressed by the product of the two angular winding numbers.
	If we use the standard parametrization without the absolute value~\cite{Shnir-2005-book}, $W=\frac{1-(-1)^{W_\theta}}{2}W_\phi$.
	The reason is that the polar angle $\theta\in[0,\pi]$ generates alternating signs of $m_z$ that unwinds each other, while the absolute sign on either $m_x$ or $m_y$ reverses the sign of the in-plane winding wherever $m_z$ unwinds in the polar direction.
	
	In the band-theory language, the above Dirac theory works at the vicinities of the Dirac points in the band structures.
	The 8-by-8 Dirac equation can be constructed by coupling the two 4-by-4 3D Dirac points.
	The three mass terms mean three independent ways or perturbations to open the bandgap.
	In condensed matter, the same low-energy Hamiltonian describes the 3D Majorana zero modes~\cite{Teo-2010-majorana,Nishida-2010-topological}, although there has been no clue where to find them.
	In topological classifications, the kink, vortex, monopole zero modes in 1D, 2D and 3D, localized at the topological defects of Dirac masses, all belong to the same Altland-Zirnbauer symmetry class BDI with integer invariant~($\mathbb{Z}$) under time-reversal and chiral symmetries~\cite{Teo-2010-topological}.
	Of course, the chiral symmetry does not rigorously present in our acoustic system, since the spectrum is not exactly up-down symmetric with respect to the Dirac frequency.
	
	\begin{figure*}[t]
		\includegraphics[scale=1.0]{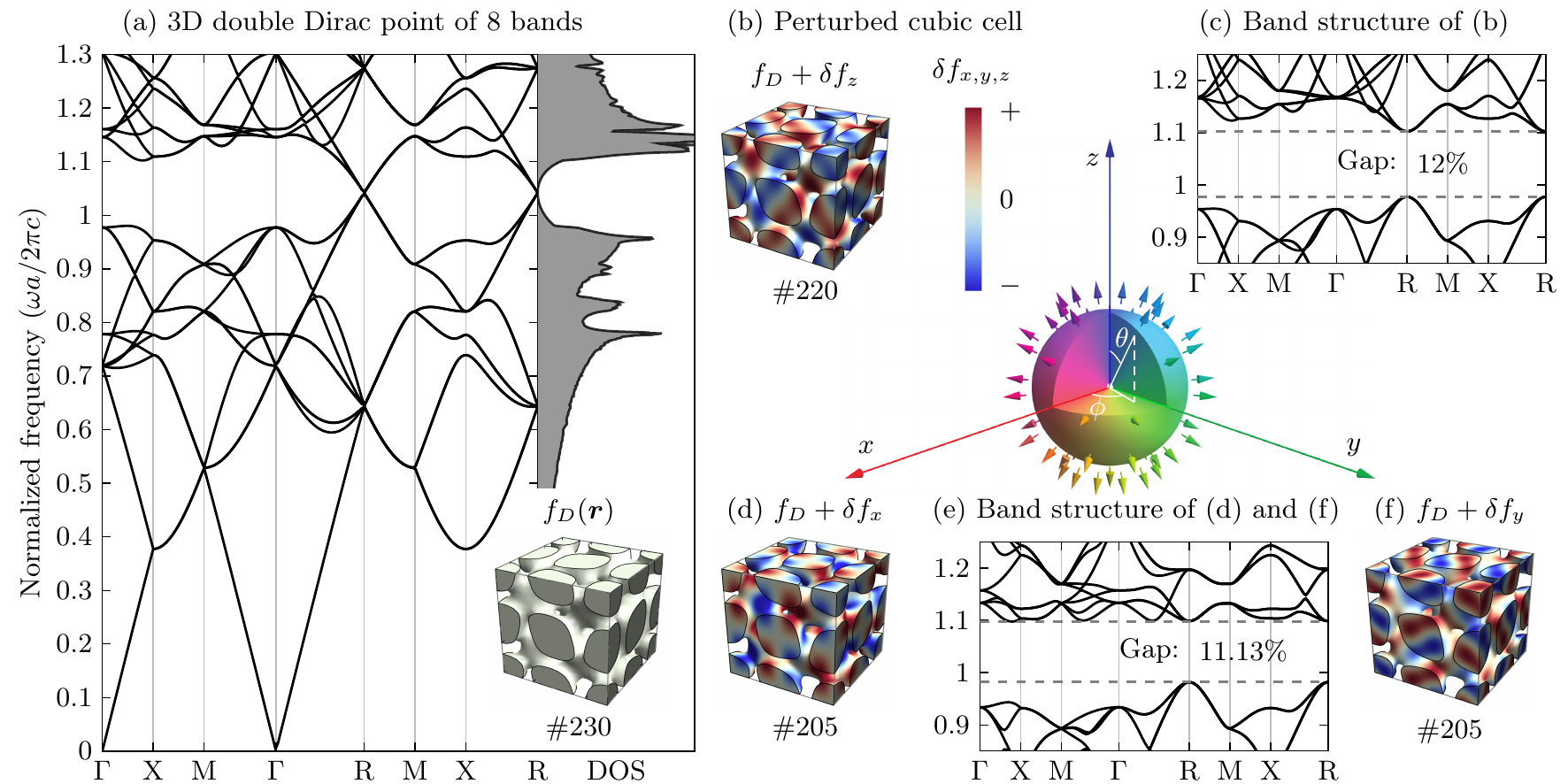}
		\caption{Acoustic realization of the double Dirac point and the three mass terms for the monopole design.
			(a)~The ideal Dirac acoustic crystal with the 8-band Dirac point in the super cell.
			(b),(d),(f)~The perturbed geometries defined by $f_D(\vecb{r})+\delta f_i(\vecb{r}), (i=x,y,z)$, whose band structures are plotted in (c) and (e).}
		\label{fig:band}
	\end{figure*}
	
	\section{Double 3D Dirac points}
	The starting point of our design is a crystal hosting two 3D Dirac points to form an 8-by-8 Dirac theory. 
	Our Dirac acoustic crystal is defined by the 3D periodic implicit functions
	\begin{equation} \begin{aligned}
			f_D(\vecb{r})&=\cycsum \sin(\frac{4\pi}{a}x)\cos(\frac{2\pi}{a}y)\sin(\frac{2\pi}{a}z)\\
			&+3\cycsum \cos(\frac{4\pi}{a}x)\cos(\frac{4\pi}{a}y),
		\end{aligned} \label{eq:230} \end{equation}
	where $a$ is the lattice constant and $\cycsum$ is the sum over the cyclic permutation of $(x,y,z)$. The volume of $f_D(\vecb{r})\ge -1.56$ is filled with sound hard material like plastic and the rest volume is  air.
	This lattice geometry, of space group $Ia\bar{3}d$~($\#230$), is improved upon a previous blue-phase-I~(BPI) structure~\cite{Cheng-2020-discovering,Lu-2016-symmetry,Cai-2020-symmetry} to have two fully frequency-isolated Dirac points, locating at the high-symmetry momenta~($\pm$P) in the Brillouin zon of the body-centered-cubic~(BCC) primitive cell.	
	By taking the simple-cubic super-cell as shown in Fig.~\ref{fig:band}, two Dirac points fold to  the momenta ${\rm R}$ point of the simple-cubic Brillouin zone, forming an 8-by-8 double Dirac point.
	The corresponding low-energy Dirac Hamiltonian is $ k_x\Gamma_{331}+k_y\Gamma_{333}+k_z\Gamma_{310}$ as listed in Fig.~\ref{fig:symmary}.
	
	Here we explain a general approach to construct photonic and phononic crystals of a specific space group $G$ using triply periodic functions $F_{G}(\vecb{r})$.
	This is done through a generic Fourier expansion~\cite{Wohlgemuth-2001-triply} 
	$F_{G}(\vecb{r})=\sum_{i}[A_i\cos(\vecb{k}_i\cdot\vecb{r})+B_i\sin(\vecb{k}_i\cdot\vecb{r})]$ with the real coefficients $A_i$ and $B_i$, satisfying $F_G(\vecb{r})=F_G(g\vecb{r})$ for each space-group element $g\in G$. The reciprocal lattice vectors $\vecb{k}=(l\frac{2\pi}{a},m\frac{2\pi}{b},n\frac{2\pi}{c})$ are defined by the lattice constant $a,b,c$ and integers $l,m,n$.
	One usually takes the lowest $l,m,n$ values for the simplest structures.
	We use this method to obtain the Dirac structure of Eq.~\ref{eq:230} and the symmetry-breaking perturbations~(masses) of Eq.~\ref{eq:230sub} to engineer the Dirac masses.
	
	\begin{figure*}[!ht]
		\centering
		\includegraphics[scale=1.0]{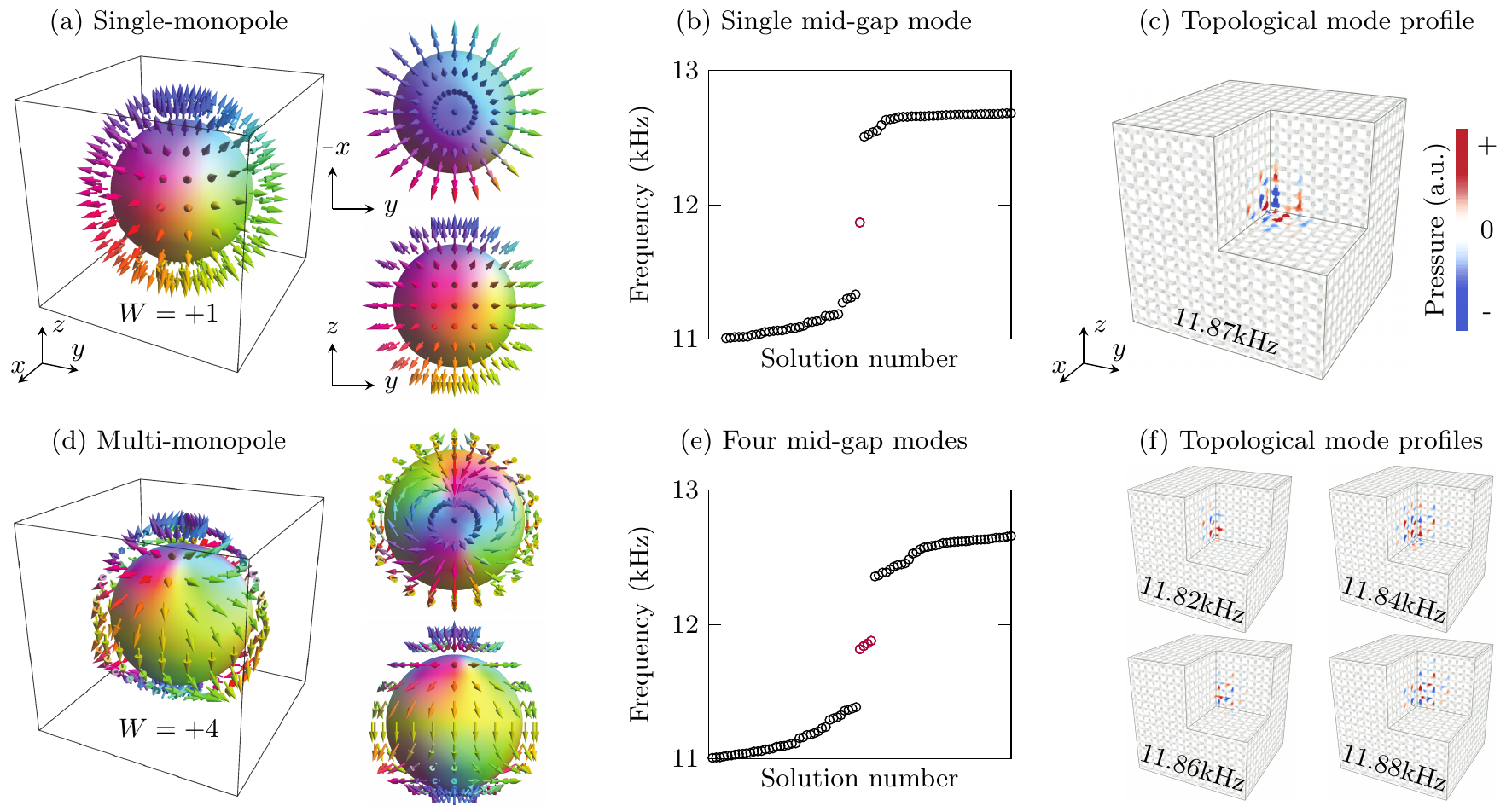}
		\caption{Single and multi-monopole resonators.
			(a), (b), (c) Dirac-mass field, eigenvalue spectrum, and the modal field distribution of the single-monopole cavity.
			(d), (e), (f) Dirac-mass field, eigenvalue spectrum, and the modal field distributions of the multi-monopole cavity with charge four~($W_\theta=W_\phi=+2$).}
		\label{fig:mode}
	\end{figure*}
	\section{Dirac-mass engineering}
	In order to build the monopole Dirac-mass field, we need three independent masses out of the total number of four mass terms: $m_x\Gamma_{100}$, $m_y\Gamma_{200}$, $m_z\Gamma_{320}$, and $m'\Gamma_{332}$ as listed in Fig.~\ref{fig:symmary}.
	The latter two are primitive-cell masses that gaps the individual Dirac point, while the first two are super-cell masses that couple the two Dirac points.
	All four masses are time-reversal symmetric, while only the first mass~($m_x$) are inversion symmetric.
	Through space-group symmetry analysis, we realize all four mass terms by perturbing the improved-BPI acoustic crystal from space group \#230 to its subgroups.
	
	The two primitive-cell masses, $m_z$ and $m'$, are the non-centrosymmetric masses for the individual Dirac point~(4-by-4) within the BCC primitive cell. 
	Since \#230 has only three maximal BCC subgroups, the non-centrosymmetric \#220, \#214~(removing inversion) and the centrosymmetric \#206~(removing a glide), we conclude that the two primitive-cell mass terms can be realized in \#220 and \#214 respectively.
	
	The two super-cell masses, the centrosymmetric $m_x$ and the non-centrosymmetric $m_y$, couple the two Dirac points in the simple-cubic super-cell. The only maximal subgroup of \#230, having both a simple-cubic lattice and inversion, is \#205~(also a maximal subgroup of \#206), reduced from \#230 by removing a primitive translation and a glide~($\{M_{01\bar{1}}|(\frac{1}{4},\frac{1}{4},\frac{1}{4})\}$), where $M_{01\bar{1}}$ is a mirror operation. Although containing inversion, \#205 inherit only half of the inversion centers from \#230, in which the two sets of inversion centers are originally connected by the broken glide.
	Consequently, there are two choices for the symmetry breakings from \#230 to \#205, in terms of which half of the equivalent inversion centers to preserve. Importantly, these two choices of \#205 designs can be chosen to be the identical structure~(Fig.~\ref{fig:band}d and \ref{fig:band}f) related by the glide operation; they correspond exactly to the two super-cell masses~($m_x$ and $m_y$) in which only one of them is inversion-symmetric.
	Conceptually, two independent mass terms sharing a symmetry-related identical structure can be understood by considering that two independent masses, forming a complex mass, are related by a phase shift associated with the translation~(in the glide). A mass term picking up a phase, through a translation, also occurs in the low-dimensional examples of the SSH model, Dirac-vortex cavity, and one-way fiber~\cite{Lu-2018-topological}.
	
	According to the space-group analysis of the four Dirac masses, we realize all of them by imposing the geometric perturbations to the original Dirac-point structure~($f_D$).
	These perturbations~($\delta f_x$, $\delta f_y$, $\delta f_z$, $\delta f'$) correspond to the Dirac masses~($m_x$, $m_y$, $m_z$, $m'$) and the required space groups~(\#205, \#205, \#220, \#214), respectively.
	\begin{equation} \begin{aligned}
			\delta f_x(\vecb{r})&=0.60 \cycsum \cos(\frac{4\pi}{a}x)\sin(\frac{4\pi}{a}y)\sin(\frac{2\pi}{a}z)\\
			\delta f_y(\vecb{r})&=0.60 \cycsum \sin(\frac{4\pi}{a}x)\cos(\frac{4\pi}{a}y)\cos(\frac{2\pi}{a}z)\\
			\delta f_z(\vecb{r})&=0.36 \cycsum \cos(\frac{4\pi}{a}x)\sin(\frac{2\pi}{a}y)\cos(\frac{2\pi}{a}z)
		\end{aligned} \label{eq:230sub} \end{equation}
	The functional forms of the first three perturbations are listed in Eq.~\ref{eq:230sub}, in which the coefficients~(amplitudes) are chosen to maximize the size of the common gap as plotted in Fig.~\ref{fig:band}.
	It is easy to check that $\delta f_y(\vecb{r})$ and $\delta f_x(\vecb{r})$ are related by the glide symmetry $\{M_{01\bar{1}}|(\frac{1}{4},\frac{1}{4},\frac{1}{4})\}$ and share the same band structure in Fig.~\ref{fig:band}~(e).
	Since $\delta f'$ opens the smallest bandgap, compared with the other three masses, we do not include it for the cavity construction.
	
	\begin{figure*}[ht]
		\centering
		\includegraphics[scale=1.0]{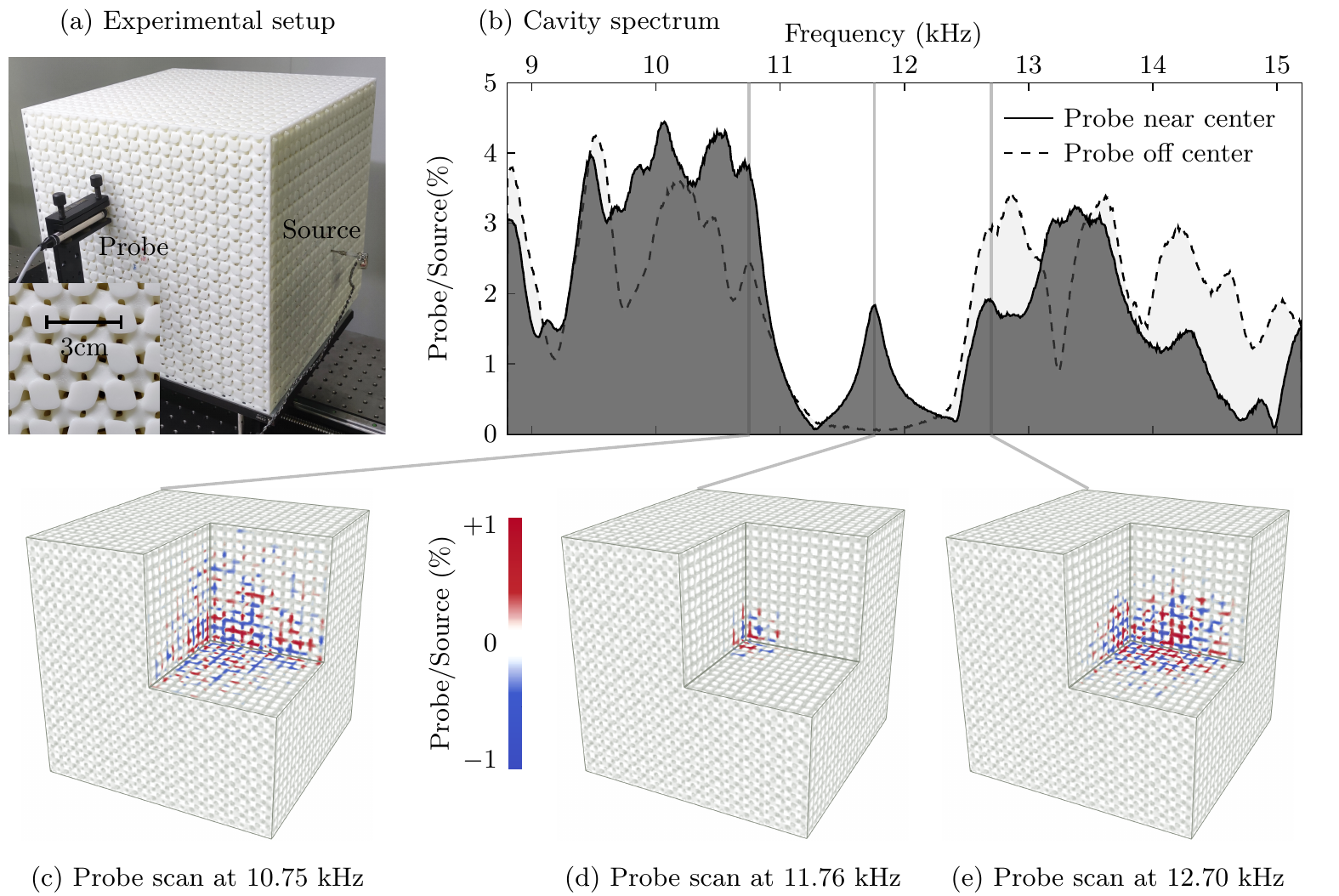}
		\caption{Experimental results on the monopole resonator. (a)~Photograph of experiment setup and a zoom-in of sample surface.
			The 3D-printed cavity has a lattice constant $a$=3 cm and a sample size of (12$a$)$^3$.
			(b)~The frequency response of the cavity showing a single mid-gap resonance.
			(c), (d), (f)~Probe scan of the pressure field distributions inside the sample at three resonant peaks across the bandgap.}
		\label{fig:exp}
	\end{figure*}
	\section{Monopole-resonator construction}
	
	With three independent mass realizations at hand, we construct the monopole cavity by applying the direction-dependent perturbations to the otherwise uniform 3D Dirac crystal, where ($\delta f_x$, $\delta f_y$, $\delta f_z$) are aligned to the three spatial axes.
	The geometry of the monopole resonator is defined by:
	\begin{equation}
		f_D(\vecb{r})+\sum_{i=x,y,z}\hat{m_i}(\vecb{r},W_\theta,W_\phi)\delta f_i(\vecb{r})\ge -1.56
		\label{func:cavity}
	\end{equation}
	where the hedgehoge mass vector $\hat{m_i}(\vecb{r},W_\theta,W_\phi)$ is given in equation~(\ref{eq:monopole}) with the monopole charge $W=W_\theta W_\phi$.
	
	To design the single-monopole cavity, we take $W_\theta=W_\phi=+1$, whose mass field is illustrated in Fig.~\ref{fig:mode}(a).
	A single mid-gap mode appears in the numerical spectrum in Fig.~\ref{fig:mode}(b), whose frequency deviation from that of the original Dirac point is less than 1\%.
	Plotted in Fig.~\ref{fig:mode}(c) is the pressure fields of the topological mode localized at the cavity center, with four cubic cells cladded in each direction.
	
	To design the multi-monopole cavity, we take $W_\theta=W_\phi=+2$ with $W=+4$ as an example.
	The high-charge hedgehog field is illustrate in Fig.~\ref{fig:mode}(d). The numerical results in Fig.~\ref{fig:mode}(e) and (f) show four nearly-degenerate mid-gap modes, whose frequency differences are within 1\%.
	
	\section{Acoustic experiments}
	
	Encouraged by the simulation results, we 3D-print the acoustic monopole cavity with stereolithography using photo-curable resin.
	Shown in Fig.~\ref{fig:exp}(a), the sample has a total volume of (36~cm)$^3$ and a lattice constant of $a$=3~cm.
	This bi-continuous structure has two types of through holes, separated by 15~mm~($\frac{a}{2}$), whose diameters are about 1.9~mm and 2.8~mm
	allowing stain-steel tubes to get inside the sample for spectral and spatial measurements.
	The sound source is an earphone whose output tube tip, diameter of 2.4~mm, is placed near the cavity center. A micro-tube probe~(B$\&$K-4182) of diameter 1.2~mm is used to detect the sound pressure any point in the sample. 
	The acoustic signal is a broadband pulse of a 12.0~kHz central frequency and a 6.4~kHz bandwidth, generated and recorded by a signal analysis module~(B$\&$K-3160-A-042), the spectrum recorded in each measurement is 100 times averaged.
	
	The two spectra in Fig.~\ref{fig:exp}(b) demonstrate the existence of the bandgap and the localized mid-gap mode. When the probe is off center, the low signal response indicates a bandgap region roughly form 11~kHz to 12.5~kHz. When the probe is centered, a resonate mode peaks in the middle of the gap at 11.76~kHz, whose quality factor is $\sim$73.
	The field profiles of the resonant modes are mapped out by scanning the probe tube inside the through holes of the sample using a motorized stage. The pressure fields are recorded, with the step size of 7.5~mm~($\frac{a}{4}$), on the three orthogonal surfaces crossing the cavity center.	In Fig.~\ref{fig:exp}(c),(d),(e), we plot the signal profiles at the frequencies of the three central resonate modes.	It is obvious that only the mid-gap mode is localized as a defect state.
	
	\section{Optimal single-mode behavior}
	
	The free spectral range~(FSR), of a resonator, is the frequency spacing between adjacent modes that shrinks as the mode volume~($V$) expands.
	A wider FSR improves the wavelength-multiplexing bandwidth, the spontaneous emission rate, and the stability of single-mode operations.
	Although conventional cavities have an inverse proportionality between modal spacing and modal volume~(FSR~$\propto V^{-1}$), the recent Dirac-vortex cavity~\cite{Gao-2020-Dirac} exhibits FSR~$\propto V^{-\frac{1}{2}}$ that drops much slower than the common scaling rule and is ideal for broad-area single-mode lasers~\cite{Yang-2022-topological}.
	Here we analyze the FSR-$V$ scaling and show that the monopole cavity of FSR~$\propto V^{-\frac{1}{3}}$ is the optimal.
	
	The FSR is fundamentally related to the density of states~${\rm DOS}(\omega_0)$, the number of states in a volume $V$ in a frequency range at the cavity resonance $\omega_0$.
	By definition, FSR is the frequency interval across which one extra mode emerges. Since ${\rm FSR}/\omega_0 \ll 1$ for large cavities, the function ${\rm DOS}(\omega_0+{\rm FSR})$ can be expanded around $\omega_0$ in Taylor series.
	\begin{equation}	\begin{aligned} 
			&{\rm DOS}\cdot {\rm FSR}\cdot V = 1\\
			&{\rm DOS}(\omega_0+{\rm FSR}) 
			\propto  D_0+D^{'}\cdot{\rm FSR}+D^{''}\cdot{\rm FSR}^2\dots
		\end{aligned} \label{eq:FSRDOS}
	\end{equation}
	where the $D_0$, $D^{'}$, $D^{''}$ are the Taylor coefficients.
	Regular optical media, such as air or uniform dielectrics, has a finite DOS with $D_0\ne 0$.
	This is why FSR~$\propto (D_0\cdot V)^{-1}$ for traditional cavities, such as the Fabry-Perot and whispering-gallery resonators.
	We note that the scaling power can be lower than -1 if the DOS diverges at a van Hove singularity in photonic crystals, but narrower FSR is not of interest for this discussion.
	
	To beat the inverse scaling law between FSR and $V$, the cavity mode has to operate at the zero DOS frequency with $D_0 = 0$, at certain (frequency-isolated) bandedge or degeneracy points.
	For example, the DOS at the linear point degeneracy is DOS$\propto(\omega-\omega_0)^{\rm{dimension}-1}$ in each dimension.
	Depending on the  asymptotic behavior, there are three different cases.
	1) $D_0 = 0$ and $D' \ne 0$, so that FSR~$\propto (D'\cdot V)^{-\frac{1}{2}}$. This is the scaling law for 2D Dirac points where the DOS grows linearly with frequency~\cite{Chua-2014-larger}.
	2) $D_0 = D' = 0$ and $D'' \ne 0$, so that FSR~$\propto (D''\cdot V)^{-\frac{1}{3}}$. This is the scaling law for 3D Dirac or Weyl points where the dispersions are all linear and the DOS grows quadratically with frequency.
	Since the linear dispersion has the lowest DOS in band theory (sublinear band dispersions result in the unphysical divergent group velocities), a further lowering of DOS would mean the bandgap opening in 3D.
	3) $D_0 = D' = D'' = 0$, which is the case of a full bandgap in which no mode exists and all Fourier coefficients of $D(\omega_0)$ vanish.
	Within the bandgap, however, defect modes can exist whose wavefunctions decay in space.
	
	\begin{figure}[t]
		\centering
		\includegraphics[scale=1.0]{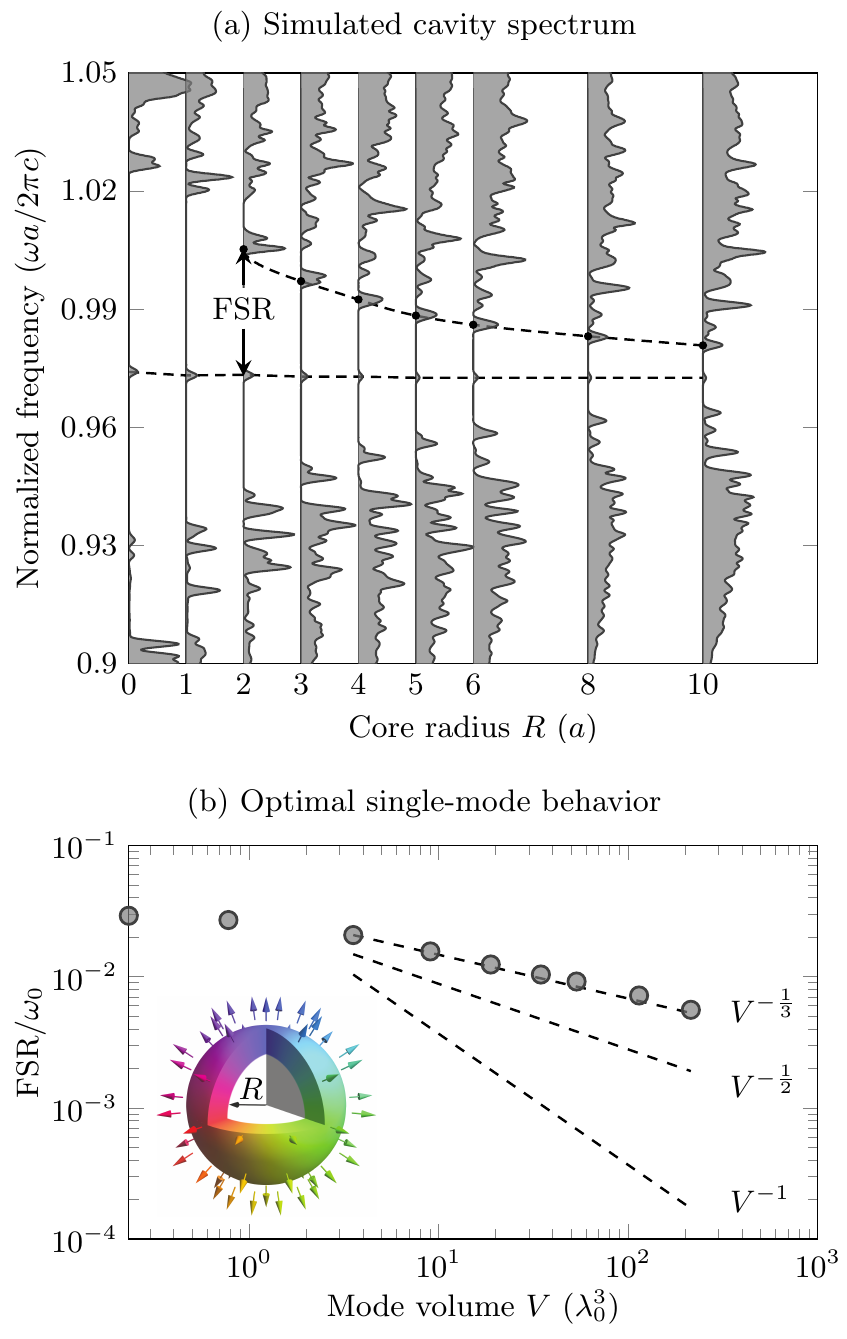}
		\caption{Numerical demonstration of the optimal FSR $\propto V^{-\frac{1}{3}}$ scaling of the monopole mode at large volumes. (a)~Resonance spectra from the cavities of core radius from 0 to 10$a$. The frequency of the monopole mode remains almost a constant at the middle of the bandgaps. (b)~Free spectral range versus mode volume relation fitted with the various scaling powers. The inset illustrates the cavity construction with the monopole core radius $R$.}
		\label{fig:fsrv}
	\end{figure}
	
	
	This DOS analysis is valid for modes of non-decaying wavefunctions, such as the plane wave or Bloch wave, whose modal volume is roughly the cavity size confined by the boundary termination. 
	Interestingly enough, the above analysis for FSR-$V$ scaling is equally valid for the defect modes of decaying wavefunctions.
	Because, in the large mode limit, the defect structure disappears into the uniform lattice, as the decay length of the defect mode diverges, and the bandedge~(or degeneracy) restores.
	For example, the quadratic bandedge restores when the depth of the quantum well diminishes and the bound state delocalize.
	Similarly, the Dirac degeneracy restores when the gap size vanishes and the mid-gap mode evolves into the Bloch mode.
	Therefore, the large-mode limit of the defect state of decaying wavefunction is precisely the bandedge state of non-decaying wavefunction, where our DOS analysis applies.
	In summary, the lowest scaling power is always achieved at the linear-dispersion points in each dimension and the best power is $-\frac{1}{3}$ in 3D.
	
	We demonstrate numerically, in Fig.~\ref{fig:fsrv}, the optimal scaling of the monopole cavity. There are two methods to enlarge the volume of the monopole mode. One method is to lower the bandgap size~\cite{Yang-2022-topological}, that is twice the FSR for the mid-gap mode. The other method is to expand the central singular point~\cite{Gao-2020-Dirac}~($R=0$) into a volume of unmodulated Dirac lattice with a finite radius $R$, shown in Fig.~\ref{fig:fsrv}(b) inset.
	As $R$ increases, the volume of the mid-gap mode increases and the high-order cavity modes drop into the bandgap.
	We take this finite-$R$ method, because it requires smaller computational domains and resources.
	For the same mode volume, the finite-$R$ cavity is more spatially confined, than the $R=0$ version, due to the non-decaying wavefunction in the gapless core and the fastest decaying tails in the cladding region by choosing the largest bandgap.
	The FSR~$\propto V^{-\frac{1}{3}}$ relation is clearly identified with the simulation results from $R=2a$ to $10a$ in Fig.~\ref{fig:fsrv}(b). 
	
	We argue that the monopole resonator is the best design for the large-volume single-mode behavior.
	Based on the above discussion, it is clear that any cavity of a spectrum converging to a 3D linear dispersion point, in the large cavity limit, shares the optimal scaling of FSR~$= (D''\cdot V)^{-\frac{1}{3}}$, including the finite-sized periodic Dirac/Weyl crystals.
	However, the non-topological approaches cannot stabilize a single mode at the center of the Dirac spectrum in the first place.
	The unique advantage of the monopole cavity is the robust mechanism in pinning a single defect mode at the middle of the bandgap and the gap size can be tuned to zero in a well-controlled fashion.

	\section{Conclusion}
	We experimentally realize the first topological mode in a 3D topological defect, completing the kink, vortex, monopole zero modes in each dimension. The next dimension ahead would be the instanton~\cite{BELAVIN-1975-instanton} in 4D space-time.
	Our approach of constructing the hedgehoges applies to other 3D Dirac/Weyl systems~\cite{Armitage-2018-weyl,Peri-2019-axial,Yang-2019-realization,Ma-2019-topological,Xia-2022-experimental,Luo2021Observation}.
	The optimal single-mode behavior of the monopole resonator could enable, in microwave and terahertz, higher-power single-mode masers or lasers.
	
	\bibliography{references}
	\bibliographystyle{unsrt}
\end{document}